\documentclass{article}

\usepackage{microtype}

    \PassOptionsToPackage{numbers, compress}{natbib}

\usepackage[main, final]{neurips_2026}

\usepackage[utf8]{inputenc} 
\usepackage[T1]{fontenc}    
\usepackage{hyperref}       
\usepackage{url}            
\usepackage{booktabs}       
\usepackage{amsfonts}       
\usepackage{nicefrac}       
\usepackage{microtype}      
\usepackage{xcolor,colortbl}         
\usepackage{graphicx}
\usepackage{amsmath}
\usepackage{cleveref}

\usepackage{wrapfig}

\newtheorem{definition}{Definition}
\newtheorem{proposition}{Proposition}
\newtheorem{corollary}{Corollary}

\newtheorem{theorem}{Theorem}

\newtheorem{proof}{Proof}

\DeclareMathOperator*{\argmax}{arg\,max}

\newcounter{axiomnum}   
\title{DeepVoting: Learning and Improving Voting Rules with Fine-Tuning}

\author{%
  Leonardo Matone\\
  Tulane University\\
  \texttt{lmatone@tulane.edu} \\
  \And
  Ben Abramowitz\\
  Tulane University\\
  \texttt{babramow@tulane.edu} \\
  \And
  Ben Armstrong\\
  Tulane University\\
  \texttt{barmstrong@tulane.edu} \\
  \And
  Avinash Balakrishnan\\
  IBM Research\\
  \texttt{avinash.bala@us.ibm.com} \\
  \And
  Nicholas Mattei\\
  Tulane University\\
  \texttt{nmattei@tulane.edu} \\
}

\begin{document}

\maketitle

\begin{abstract}
Recently, social choice theory has been found broadly useful for AI alignment and evaluation, often based on the axiomatic properties for combining preferences provided by social choice functions. However, for many sets of axioms, functions which \textit{never} violate the axioms are known to not exist, e.g., Arrow's Impossibility Theorem. Recent work on learning novel rules to \textit{minimize} axiom violations shows the effectiveness of machine learning but explores a limited class of axioms.
In this work we show the effectiveness of standard social choice data structures as features for learning existing voting rules, resulting in networks that have better accuracy than past approaches while using an order of magnitude fewer parameters. Subsequently, we develop a process for \textit{adding axiomatic properties} to existing voting rules. By building novel axiomatic loss functions we are able to use transfer learning to fine-tune existing voting rules in a way that quantitatively preserves their original behaviour while also exhibiting strong adherence to previously absent axiomatic properties, resulting in novel voting rules that are \emph{closer} to the impossibility frontier than those in the machine learning or theoretical literature.

\end{abstract}

\section{Introduction}
\label{sec:introduction}

Preference aggregation is an important and widespread problem across many areas of computer science including information retrieval, recommender systems, reinforcement learning with human feedback, and agent evaluation. A common goal when choosing a preference aggregation function is that of axiom satisfaction: finding functions which \textit{guarantee} certain mathematical properties. Unfortunately, dozens of ``impossibility theorems'' have shown that many sets of axioms cannot be mutually satisfied~\cite{brandt2016handbook}.
In this work we provide the foundational theoretical results, previously absent, required for a scientifically rigorous approach to learning voting rules. We demonstrate that novel rules, learned from base models trained to replicate existing well-understood rules, can be imbued with axiomatic properties not originally present. Our results highlight that in the learning of social choice axioms \textit{some impossibilities are harder than others}.

Recently, a growing body of work has shown significant applicability of social choice theory to AI training \cite{dai2024mapping}, alignment \cite{procaccia2025clone,heymann2025adaptive,maura2025jackpot} and evaluation \cite{lanctot2023evaluating,lanctot2026active}. Learning social choice mechanisms can allow us to efficiently search for novel functions which provide desirable properties. Past work often shows that no single voting rule can perfectly align AI systems \cite{mishra2023ai,siththaranjan2023distributional}, suggesting a demand for novel, tailored social choice functions. Other work has shown that probabilistic functions, i.e., lotteries, can provide stronger axiomatic performance for alignment settings \cite{maura2025jackpot}.

This work provides a principled and rigorous methodology for learning both existing and novel voting rules. By using carefully chosen input embeddings we overcome the limitations of much past work, which has been limited by large networks with complex architectures~\cite{anil2021learning}, limited data~\cite{burka2022voting}, and the number of voters that can be considered simultaneously~\cite{anil2021learning,hornischer2025learning}. We demonstrate that, across a wide range of voting rules, \textit{we are able to train models which replicate existing rules better than the previous state-of-the-art with significantly fewer parameters}.

The majority of our results relate to axiomatic social choice. Axioms in social choice theory fall into one of two categories: intra-profile axioms (e.g., the majority winner property) which can be said to have been violated, or not, based on a social choice function and a single preference profile~\cite{fishburn2015theory,abramowitz2025axiomatic} and inter-profile axioms which require more information. For example, the Participation property is violated if a voter can improve their outcome by not participating. To determine whether this occurs it is necessary to consider both the original profile and the profile in which the voter abstains~\cite{schmidtlein2022voting}.

While many axioms are intra-profile, many important axioms require evaluation of multiple alternative profiles, such as Participation and Strategyproofness. Little past work has explored the ability of machine learning to develop learned rules that minimize violations of inter-profile axioms~\cite{hornischer2025learning}. 
Building upon base models trained to replicate existing rules, \textbf{our primary contribution is a fine-tuning approach which effectively adds new axiomatic properties into existing voting rules} by developing custom loss functions based on the principles of stochastic dominance. The end result of the process is a voting rule (in the form of a neural network) that behaves quantitatively similarly to an existing voting rule while exhibiting a novel axiomatic property.


Experimental studies of learned voting rules face two primary critiques: the relative complexity of using a neural network in place of a simple function, and reduced interpretability from the black-box nature of machine learning models. Both of these concerns are valid but tangential to our work; we aim first to provide a novel rigorous exploration of the data best suited for learning existing voting rules and then to develop novel voting rules with superior axiomatic properties to those known in the literature. A lack of interpretability is acceptable for us as our focus is on understanding the possibilities of learning, rather than deploying a rule for widespread use in human settings. However, in non-political settings that utilize social choice such as AI alignment or recommender systems, where deployed systems often aggregate orders using bandits or other learned, ``black box'' procedures \cite{aird2024dynamic,patro2020fairrec}, the lack of simple, explainable rules is often a secondary concern to system efficiency.


\paragraph{Contribution.}
We (1) demonstrate, for the first time, that using common embeddings from social choice dramatically reduces the complexity and increases the efficiency of learning voting rules, and (2) use transfer learning (fine-tuning) to add axiomatic properties to learned rules. Our results are broadly useful for work using machine learning to develop voting rules and specifically useful as a learned probabilistic social choice function (i.e., a lottery) used in many modern AI evaluation schemes and recommendation systems where the increased efficiency from randomization is desirable.

\section{Related Work}

Machine learning has been applied to many tasks adjacent to voting, including auction design, matching, allocation mechanisms, and learning complex preference structures~\cite{dutting2019machine,pavlov2011optimal,malakhov2008optimal,curry2022learning,peri2021preferencenet,ravindranath2021deep}.

While formal proposals to learn voting rules date back over a decade~\cite{procaccia2009learnability,xia2013designing}, attention to learning voting rules has increased considerably in recent years.
\citet{kujawska2020predicting} and \citet{burka2022voting} used several common machine learning methods to mimic existing voting rules. However, both of these works overlooked the importance of the choice of embedding in the role of learning, finding that certain rules were ``easier'' to learn but not theoretically characterizing why certain embeddings maintain properties. 
Subsequently, \citet{anil2021learning} showed that permutation-invariant architectures offer better generalization to larger numbers of voters. \citet{hornischer2025learning} consider CNNs and additional architectures, and \citet{li2026learning} use graph neural networks. These papers continue to have a primary focus on learning existing voting rules and do not consider the ability of well-chosen embeddings to improve outcomes. 
While \citet{armstrong2025power} considered the impact of embeddings on rule learnability across a wide range of preference distributions the results were limited to only 5 candidates and provided little subsequent analysis.




\citet{procaccia2009learnability} showed that positional scoring rules are efficiently PAC learnable, but learning pairwise comparison-based voting rules requires an exponential number of samples. 
\citet{firebanks2020machine} uses one axiomatic measure of optimality (Condorcet consistency) and strategyproofness for learning, but as we show, their choice of embedding does not preserve the information necessary to implement Condorcet-consistent rules, hindering performance. \citet{wilson2019generating} focused on learning a voting rule given pair-wise relations and properties that must hold for the optimization criteria. However, they examine only the possibility of learning and do not employ any ML techniques.

The loss function chosen by~\citet{armstrong2019machine} was a function of the profile and outcome, and thus could learn a rule but not inter-profile axioms such as Participation. 
Recently, \citet{mohsin2022learning} focused on the problem of designing and/or learning fair and private rules using random forests and a subset of embeddings we study, proving that under differential privacy there is an upper bound on the trade-off between group fairness and efficiency.
\citet{caiata2026what} consider a wide range of intra-profile axioms in a multi-winner setting. They learn novel rules which violate axioms less than existing rules, but their methodology does not transfer to inter-profile axioms or finetuning of existing rules.
Learning voting rules bears some similarity to the well studied area of learning to rank (L2R) from the machine learning literature \cite{cao2007learning}. 
L2R is concerned with accurate recovery of the \emph{population preference} and not the axioms or properties of the aggregation method itself (e.g., fairness).

\section{Preliminaries}
\label{sec:preliminaries}

\paragraph{Agents and Preference Profiles}
Let $V$ be a set of $n$ voters and $C$ a set of $m$ candidates.
Each voter $i \in V$ reports a strict order $x_i$ over all candidates in $C$ as their ballot, i.e. a permutation of the candidates in $C$.
We denote that $i$ strictly prefers $a$ over $b$ by $a \succ_i b$ for $a,b \in C$.
There are $m!$ possible ballots, or ways to strictly order (permute) the candidates in $C$.
A list of $n$ ballots, one for each voter, constitutes a \emph{profile} $X = (x_i)_{i \in V}$.
Voter $i$ ranks candidate $a$ at position $1 \leq x_i(a) \leq m$, and we use $x_i[l]$ to denote the $l^{th}$ element of $x_i$.
Let $\mathcal{X}$ be the set of all possible profiles.

\paragraph{Probabilistic Social Choice Functions}
A probabilistic social choice function (PSCF) is a function $f : \mathcal{X} \rightarrow \Delta(C)$ that takes a profile $X \in \mathcal{X}$ as input and returns a \emph{lottery}, or probability distribution $f(X) \in \Delta(C)$ over the set of candidates in the profile, where $\Delta(C)$ is the set of all lotteries over $C$. 
Let $\mathcal{F}$ be the set of all such PSCFs.
We denote by $f(X)(a)$ the probability mass that $f$ places on candidate $a$ given profile $X$.
Any PSCF can be used to construct a non-deterministic voting rule by sampling a winner from the lottery.
%
%
The PSCFs we consider return a lottery that is a uniform distribution over a non-empty subset of the candidates, i.e., there are multiple potential winners (ties) that we choose from to construct a single-winner voting rule. Therefore, we let $U(Y)$ denote the uniform distribution over any finite set $Y$.
%
We will abuse notation and let $f(X,c)$ denote the probability mass that lottery $f(X)$ allocates to candidate $c \in C$.

\subsection{Embedding Social Preferences}
Traditional feed-forward neural networks require a fixed-size input for learning and inference, corresponding to the size of their input layer. If we learn voting rules using networks that take the entire profile as input, then the input layer must be large $(m \times n)$ and networks cannot scale to work with arbitrarily large numbers of voters.
Similarly, if the number of voters shrinks, then the network must be retrained or, as demonstrated by~\citet{hornischer2025learning}, the profile may be padded carefully to preserve performance. To learn rules that are agnostic to the number of voters we need embeddings of a fixed-size that retain relevant information for profiles with any number of voters. Naturally, different embeddings preserve different information from the original profile, leading to differing efficacy when learning different rules and axioms. Most rules and axioms in the literature are defined for any positive number of voters, so we would like a learned rule to be similarly agnostic.

\begin{figure}
    \centering
    \includegraphics[width=1\linewidth, trim=32 20 29 20,
  clip]{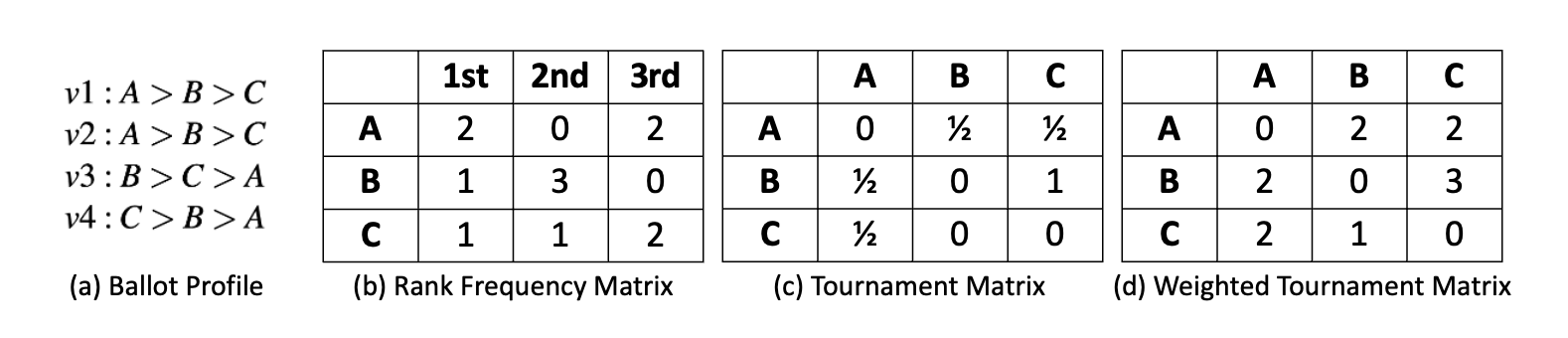}
    \caption{Three embeddings derived from the ballot profile. The size of the profile grows in $O(mn)$, while each embedding grows with $O(m^2)$, which is far smaller in the most common case, when $m \ll n$. However, these embeddings do not preserve all of the information in the original profile.}
    \label{fig:embeddings_example}
\end{figure}

An embedding $T$ is a function $T: \mathcal{X} \rightarrow \mathcal{X}'$ mapping profiles to some codomain $\mathcal{X}'$.
The embeddings we are concerned with are many-to-one mappings. This means that multiple different profiles may have the same embedding and consequently, some information about the original profile is lost in the embedding.
%
The following embeddings, illustrated in \Cref{fig:embeddings_example}, are drawn from the voting literature but are not commonly discussed as embeddings in the machine learning literature.

\begin{definition}[Tournament Embedding]
    Embedding $T_T$ yields an $m \times m$ matrix $M$ where $M[j,k] = 1$ if a majority of voters prefer $j$ over $k$, $M[j,k] = \frac{1}{2}$ if an equal number of voters prefer each candidate, and $M[j,k] = 0$ otherwise.
\end{definition}

\begin{definition}[Weighted Tournament Embedding]
    Embedding $T_{WT}$ yields an $m \times m$ matrix $M$ where $M[j,k] = |\{i \in V : j \succ_i k\}|$ for $j,k \in C$.
\end{definition}

\begin{definition}[Rank Frequency Embedding]
    Embedding $T_{RF}$ yields an $m \times m$ matrix $M$ of how many voters rank each candidate $c \in C$ in each position $k \leq m$ where $M[c,k] = |\{i \in V \text{ s.t. } x_i(c) = k\}|$.
\end{definition}

Observe that $T_{WT}$ contains strictly more information about the original profile than $T_T$ as the tournament can be computed from the weighted tournament. 
Additionally, we normalize $T_{RF}$ and $T_{WT}$ by dividing all their elements by $n$.
There is a tension in the literature between rules that use \textit{positional information} and those that rely on \textit{pairwise comparison} information~\cite{brandt2016handbook}. Note $T_{RF}$ maintains positional information while $T_{WT}$ and $T_T$ are majoritarian.
At times we also refer to a concatenation of all three embeddings which we refer to as $T_{CO}$: 

\begin{definition}[Combined Embedding]
    The combined embedding $T_{CO}$ is the $3m^2$ concatenation of $[T_{RF}, T_{T}, T_{WT}]$.
\end{definition}



\subsection{Probabilistic Social Choice Functions}

\begin{wraptable}{r}{8.7cm}
\caption{
Informational requirements of PSCFs and which axioms each function satisfies.
}
\vspace{0.7em}
\centering
\renewcommand{\arraystretch}{0.95}
\setlength{\tabcolsep}{12pt}
\addtolength{\tabcolsep}{-0.5em}
\begin{tabular}{@{}lcc|cccc@{}}
\toprule
                        & \textbf{\rotatebox{90}{Fishburn}} & \textbf{\rotatebox{90}{Scoring}} & \textbf{\rotatebox{90}{Participation}} & \textbf{\rotatebox{90}{Monotonicity}} & \textbf{\rotatebox{90}{Strategyproof}} & \textbf{\rotatebox{90}{Top Cycle}} \\ \midrule
\textbf{Plurality}      & C3                                & \checkmark                       & \checkmark                             & \checkmark                            &$\times$                       & $\times$                           \\
\textbf{Borda}          & C2                                & \checkmark                       & \checkmark                             & \checkmark                            &$\times$                       & $\times$                           \\
\textbf{Copeland}       & C1                                & $\times$                         & $\times$                               & \checkmark                            &$\times$                       & \checkmark                         \\
\textbf{Schulze}        & C2                                & $\times$                         & $\times$                               & \checkmark                            &$\times$                       & \checkmark                         \\
\textbf{Simpson-Kramer} & C2                                & $\times$                         & $\times$                               & \checkmark                            &$\times$                       & $\times$                           \\
\textbf{IRV}            & C3                                & $\times$                         & $\times$                               & $\times$                              &$\times$                       & $\times$                           \\
\textbf{Black's Rule}   & C2                                & $\times$                         & $\times$                               & \checkmark                            &$\times$                       & $\times$                           \\
\textbf{Ranked Pairs}   & C2                                & $\times$                         & $\times$                               & \checkmark                              &$\times$                       & \checkmark                         \\ \bottomrule
\end{tabular}
\label{tab:rule-descriptions}
\end{wraptable}

We consider 8 unique PSCFs through the paper, listed in \Cref{tab:rule-descriptions}. Complete definitions of each rule can be found in \Cref{app:rules}. We discuss here the information required to compute each PSCF. There exist numerous ways to categorize the informational requirements of voting rules;
\citet{fishburn1977condorcet} described three categories: C1 rules are those which can be computed using only information found in the tournament matrix $T_T$, C2 rules are those which can be computed using information found in the \textit{weighted} tournament matrix $T_{WT}$, and all other rules are C3 rules. An alternative categorization is that of \textit{positional scoring rules}, rules which can be computed using exactly the information of $T_{RF}$, the rank frequency embedding \cite{brandt2016handbook}.
Rules can also be discussed in terms of the axioms they satisfy. We consider four axioms in this work, discussed in \Cref{sec:axiom_loss_fns}, and list in \Cref{tab:rule-descriptions} which rules are known to satisfy or violate each axiom~\cite{brandt2016handbook,moulin1980strategy}.

\subsection{Axiom Loss Functions} \label{sec:axiom_loss_fns}

We consider four well-known social choice axioms -- Top Cycle, Participation, Monotonicity, and Strategyproofness -- and define novel loss functions that quantify how ``close'' a PSCF is to satisfying each axiom. In \Cref{sec:resist} we demonstrate how to fine-tune pre-trained base models by training them on a combination of axiom loss function and rule loss that move them towards satisfaction of each individual axiom while maintaining rule performance.


It is useful to first introduce adjacent swaps and stochastic dominance. An \textit{adjacent swap} refers to swapping the places of any two adjacent candidates in an ordering. With $m$ candidates there are $m-1$ possible adjacent swaps for any ordering, e.g., given $(a,b,c)$ the orderings $(b,a,c)$ and $(a,c,b)$ are the only two adjacent swaps. We denote by $s(X, i, j)$ the profile obtained by swapping candidates ranked at positions $j$ and $j+1$ in $x_i$. 
\textit{Stochastic dominance} (SD) is a relation between two lotteries with respect to a given ordering of the candidates. We use stochastic dominance to characterize voters' preferences over lotteries based on their preferences over candidates.

\begin{definition}[Stochastic Dominance]
    Given a strict ordering $x$ over $C$ and two lotteries $P$ and $Q$, we say $P$ stochastically dominates $Q$ with respect to $x$ if for all $k \leq m$, $\sum\limits_{l \leq k} P(x[l]) \geq \sum\limits_{l \leq k} Q(x[l])$.
\end{definition}


\begin{definition}[Stochastic Dominance Loss]
    Given ordering $x$ over $C$, a reference lottery $P$, and an alternative lottery $Q$, we say that the stochastic dominance loss is zero if $P$ stochastically dominates $Q$. If $P$ does not stochastically dominate the alternative lottery $Q$, then the loss is equal to $L^\text{sd}(P | x, Q) = \max\limits_{k \leq m} (\sum\limits_{l \leq k} Q(x[l]) - \sum\limits_{l \leq k} P(x[l]) )$, i.e. the largest difference between the sums of prefixes of the lotteries over all prefixes when the distributions' supports are ordered by $x$.
\end{definition}


We are now ready to introduce our four axioms and define the four loss functions for measuring how close a PSCF is to satisfying them with respect to a given profile.

\subsubsection{Axiom \theaxiomnum: Top Cycle}
\stepcounter{axiomnum}

The Top Cycle of a profile is the smallest set of candidates such that all candidates inside the set preferred by a weak majority of voters to all candidates outside the set ~\cite{brandt2016handbook}. The Top Cycle axiom says that all probability mass must be on candidates in the Top Cycle. We translate this to a loss function by measuring the total probability mass allocated to candidates outside of the Top Cycle.

\begin{definition}[Top Cycle Loss]
    Let $TC(X)$ refer to the set of candidates in the Top Cycle for a profile $X$ and $f$ be a PSCF, then  $L^\text{TC}(X,f) = 1 - \sum\limits_{c \in TC(X)} f(X,c)$.
\end{definition}

\subsubsection{Axiom \theaxiomnum: Monotonicity}
\stepcounter{axiomnum}

A PSCF is (weakly) monotonic if swapping the positions of any two candidates in the ordering of any single voter never decreases the probability mass on the candidate swapped upward. A PSCF is adjacent-swap monotonic (ASM) if swapping any two adjacent candidates in a single ordering within the profile will never decrease the probability mass on the candidate swapped upward. These properties are equivalent: Monotonicity clearly implies ASM, and ASM implies monotonicity as we can create a series of adjacent swaps that move one candidate up and the candidate formerly in that position down to the other candidates' original position. We characterize our loss function in terms of ASM as the space of all $n(m-1)$ possible adjacent swaps in a profile is much smaller than the space of all $\frac{nm(m-1)}{2}$ possible swaps, and we therefore expect to learn ASM more easily. We generalize ASM to consider settings where only the $k$ top candidates in a voter's ordering can be swapped. When $k = m-1$ this is equivalent to full ASM.
    A PSCF $f$ satisfies Adjacent-Swap Monotonicity if for all possible profiles $X \in \mathcal{X}$, for all voters $i \in V$, if we take any two adjacent candidates in $x_i$ where $c \succ_i c'$ and swap them, this does not decrease the probability mass allocated to $c'$. 



\begin{definition}[Adjacent-Swap Monotonicity Loss]
    Given a profile $X$ and an index $k$, $L^\text{mono}(f, X, k) = \max \big(0, \max\limits_{i \in V, j \leq k} f(X, x_i[j]) - f(s(X, i, j),x_i[j-1]) \big)$.
\end{definition}

%

\subsubsection{Axiom \theaxiomnum: Participation}
\stepcounter{axiomnum}

A PSCF satisfies Participation if no voter prefers the outcome when abstaining over casting a truthful ballot. Violation of Participation is referred to as the No Show Paradox.

We denote by $X^{-i}$ the profile obtained when voter $i$'s vote is removed from $X$, representing an abstention. We say that voter $i$ prefers voting truthfully to abstaining when $f(X)$ stochastically dominates $f(X^{-i})$ with respect to $x_i$. We want our PSCF to be immune to strategic abstentions. 
    A PSCF $f$ obeys Participation if, $\forall\ X \in \mathcal{X}$, and $\forall\ i \in V$, $f(X) \succ_i f(X^{-i})$, so every voter prefers truthful voting to abstention because $f(X)$ stochastically dominates $f(X^{-i})$ according to $x_i$. 


\begin{definition}[Participation Loss]
    Given a profile $X$ and PSCF $f$,
    $L^\text{part}(f,X) = \max\limits_{i \in V} L^\text{sd}(f(X) | x_i, f(X^{-i}))$, the maximum SD loss for any single abstaining voter.
\end{definition}

\subsubsection{Axiom \theaxiomnum: Strategyproofness}
\stepcounter{axiomnum}

Adjacent-Swap Strategyproofness (ASSP) was originally described as a means of capturing whether minor changes to a voter's preference order (``small lies'') may provide useful nonmanipulability results but has been shown to be equivalent to general forms of strategyproofness on universal preference domains~\cite{sato2013sufficient,gibbard1977manipulation}.


A PSCF satisfies ASSP if there is no profile for which a single voter $i$ can make an adjacent swap in their preferences to create an outcome that is not stochastically dominated by the truthful outcome with respect to $x_i$.
Formally, a PSCF is adjacent-swap strategyproof if $\forall\ X \in \mathcal{X}$, $\forall\ i \in V$, $\forall\ j \in 1 \dots m-1$, 
$L^\text{sd}(f(X) | x_i, f(s(X, i, j))) = 0$.
%
We generalize ASSP to consider strategyproofness in which only the $k$ top candidates in a voter's ordering can be swapped, and define a corresponding loss function. For experimental tractability we limit $k$, but when $k = m-1$ this is equivalent to full ASSP.


\begin{definition}[Adjacent-Swap Strategyproofness Loss]
Given a profile $X$ and an index $k$, $L^\text{assp}(f, X, k) = \max\limits_{i \in V,\ j \leq k} L^\text{sd}(f(X) | x_i, f(s(X, i, j)))$.
\end{definition}

\section{Social Choice Embeddings Enable SOTA Rule Learning}
\label{sec:rule_learn}

To achieve our primary goal of learning \textit{novel} rules which behave in a quantitatively similar manner to existing rules, we first show that we are able to achieve state-of-the-art performance at learning existing rules by using the input embeddings defined in \Cref{sec:preliminaries}. Using networks with a simpler structure and fewer parameters we surpass or match the performance of all past work. These networks serve as base models for our subsequent fine-tuning task.

Our embeddings provide three key advantages: (1) They dramatically reduce the size of our input layer, therefore greatly reducing the number of model weights. 
(2) When a rule is paired with an appropriate embedding, the embedding preserves all the information necessary to learn the rule. (3) Input size no longer depends on the number of voters; as a result training costs increase only with the number of candidates and a trained model can be used for profiles containing any number of voters.

\paragraph{Experimental Setup and Network Details}
We train our base models on profiles with $n=44$ voters and $m=11$ candidates. For all experiments, profiles are sampled from the \textit{Impartial Culture} distribution -- i.e., rankings are generated uniformly at random~\cite{black1958theory}. We use the Whalrus package for voting rule implementations.\footnote{https://pypi.org/project/whalrus/}

We use an MLP architecture similar to that used by \citet{anil2021learning}, with 5 fully-connected layers, the first hidden layer with 200 nodes, then four with 120 nodes, layer normalization between each hidden layer, TanH activation functions, a Softmax output layer.\footnote{Discussion of other setups is in \Cref{app:other_test}.} The key difference is that our network takes in embedded profiles so the size of our input layer is $m^2$ ($T_T$, $T_{RF}$, $T_{WT}$) or $3m^2$ ($T_{CO}$) compared to their $nm^2$. This reduces the total number of parameters in our model from 10 million to  $\approx$150K. We train our models on a set of 100,000 randomly sampled profiles in batches of size 32 for 1000 epochs. Models are trained 5 times using different seeds.
We train each model on an NVIDIA A100 GPU with each model requiring $\approx$4 hours. We used the Adam optimizer with an initial learning rate of 0.001, reducing on plateau (patience = 50, factor = 0.5, min\_lr = $1e^{-5}$).

Models receive as input one or three flattened embedding matrices. e.g., models trained on $T_{CO}$ receive: the rank frequency, tournament, and weighted tournament matrices. For each PSCF we train a base model on \textbf{L1 Loss}, the L1 distance between the output of the model and the target PSCF.

\begin{table}[tbp]
\caption{Single winner accuracy of our base models compared with prior work on rules considered in past work. Using fewer parameters we are able to nearly match, or exceed, all past results.}
\centering
\begin{tabular}{lccccc}
\hline
\textbf{}                      & \textbf{\# Parameters} & \textbf{Plurality} & \textbf{Borda} & \textbf{Copeland} & \textbf{SK}   \\ \hline
DeepVoting (ours)              & \textbf{150K}          & 0.99               & 0.95           & \textbf{1.00}     & \textbf{0.80}          \\
\citet{anil2021learning}       & 10M                    & \textbf{1.00}      & 0.99           & 0.83              & \textbf{0.80} \\
\citet{hornischer2025learning} & 1.2M - 1.8M            & 0.99               & 0.95           & 0.90              & N/A           \\
\citet{li2026learning}         & 100K-1M                & \textbf{1.00}      & \textbf{1.00}  & 0.76              & 0.73          \\ \hline
\end{tabular}
\label{tab:comparison_with_others}
\end{table}

\paragraph{Base Model Performance}

\begin{figure*}[tbp]
    \centering
    \includegraphics[width=\linewidth]{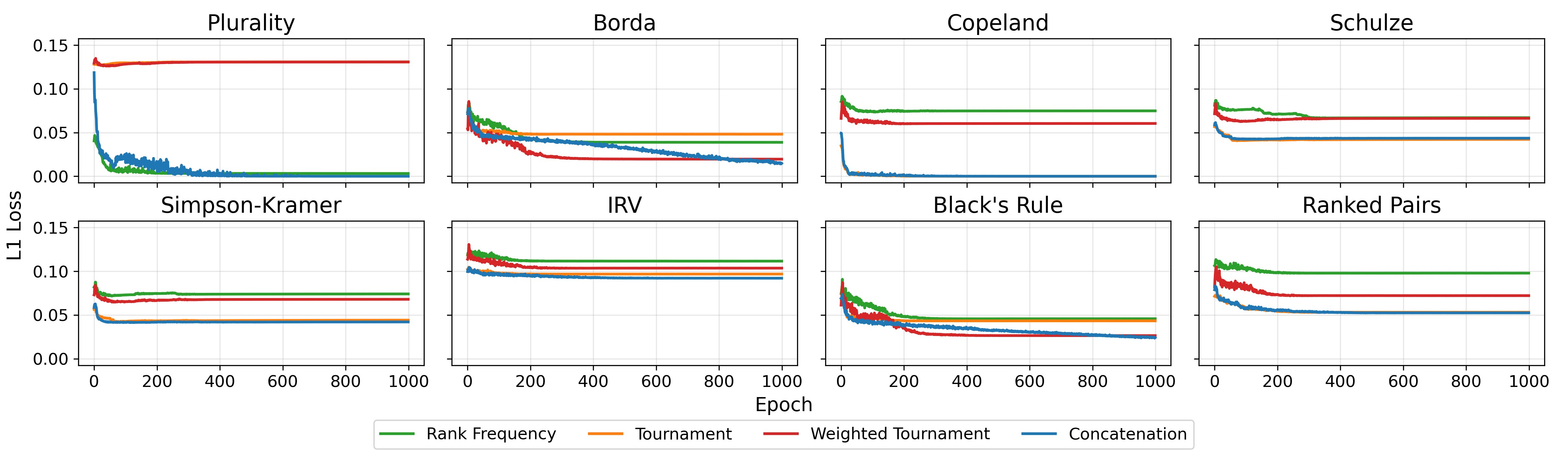}
    \caption{
    Validation Loss per epoch for each rule and embedding pair.
    }
    \label{fig:rule_validation_loss}
\end{figure*}

\Cref{tab:test_acc_all_rules_IC} gives the final validation accuracy and 
\Cref{fig:rule_validation_loss} shows plots of validation losses during training, using 10,000 samples from a held-out validation set, to demonstrate the effectiveness of learning for each rule-embedding pair.
Most rules reach their lowest loss quite quickly, with the exception of Borda's and Black's rules.
%
Among settings where the embedding does not provide sufficient information to learn a rule completely we do often see limited loss reduction (e.g., $T_T$ with Ranked Pairs). Rule loss on models trained with $T_T$ is often much lower than random, suggesting that the simple binary structure of the $T_T$ matrix is relatively easy to learn from.
On the other hand, while the $T_{WT}$ features contain strictly more information than $T_T$ they often result in much less learning than $T_{CO}$ even on C1/C2 rules which are computable from $T_{WT}$ (e.g., Copeland, Schulze, Simpson-Kramer, Ranked Pairs). In these cases the $T_{CO}$ embedding rarely outperforms $T_T$, suggesting that the network may only be learning from the simpler $T_T$ features.

\begin{table}[tbp]
\caption{
Accuracy of our models on each embedding using test data from the Impartial Culture $(m=11, n=44)$, on profiles that may contain ties. A prediction counts as correct if the alternative receiving the most predicted probability mass is a (potentially tied) winner. Cells show mean $\pm$ standard error over 5 independently trained models. Most accurate embeddings are shown in bold. Shaded cells indicate the embedding contains sufficient information to learn the rule perfectly.
}
\centering
\begin{tabular}{@{}lcccc@{}}
\toprule
\textbf{}               & $T_{RF}$ & $T_{T}$ & $T_{WT}$ & $T_{CO}$ \\ \midrule
\textbf{Plurality} & \cellcolor{green!25}\textbf{0.9887 $\pm$ 0.0254} & 0.3284 $\pm$ 0.0049 & 0.3290 $\pm$ 0.0031 & \cellcolor{green!25}0.9884 $\pm$ 0.0259 \\
\textbf{Borda} & \cellcolor{green!25}0.8168 $\pm$ 0.0618 & 0.6945 $\pm$ 0.0021 & \cellcolor{green!25}0.8945 $\pm$ 0.0387 & \cellcolor{green!25}\textbf{0.9501 $\pm$ 0.0036} \\
\textbf{Copeland} & 0.6403 $\pm$ 0.0304 & \cellcolor{green!25}0.9999 $\pm$ 0.0001 & \cellcolor{green!25}0.7184 $\pm$ 0.0094 & \cellcolor{green!25}\textbf{1.0000 $\pm$ 0.0000} \\
\textbf{Schulze} & 0.6930 $\pm$ 0.0156 & 0.8115 $\pm$ 0.0032 & \cellcolor{green!25}0.7144 $\pm$ 0.0049 & \cellcolor{green!25}\textbf{0.8119 $\pm$ 0.0083} \\
\textbf{Simpson-Kramer} & 0.6924 $\pm$ 0.0125 & 0.7969 $\pm$ 0.0195 & \cellcolor{green!25}0.7045 $\pm$ 0.0124 & \cellcolor{green!25}\textbf{0.8036 $\pm$ 0.0023} \\
\textbf{IRV} & 0.3781 $\pm$ 0.0228 & 0.3766 $\pm$ 0.0158 & 0.3622 $\pm$ 0.0139 & \textbf{0.4061 $\pm$ 0.0220} \\
\textbf{Black's Rule} & 0.7856 $\pm$ 0.0380 & 0.7099 $\pm$ 0.0021 & \cellcolor{green!25}\textbf{0.9077 $\pm$ 0.0353} & \cellcolor{green!25}0.8654 $\pm$ 0.0109 \\
\textbf{Ranked Pairs} & 0.4675 $\pm$ 0.0304 & \textbf{0.6778 $\pm$ 0.0349} & \cellcolor{green!25}0.5189 $\pm$ 0.0227 & \cellcolor{green!25}0.6561 $\pm$ 0.0497 \\ \bottomrule
\end{tabular}
\label{tab:test_acc_all_rules_IC}
\end{table}

An unexpected result in \Cref{tab:test_acc_all_rules_IC} is the limited accuracy of the Borda network trained using $T_{RF}$, despite the fact that $T_{RF}$ contains sufficient information to learn the rule. We hypothesize that this is due to the relative complexity of calculating Borda using $T_{RF}$. While Borda scores can be calculated from $T_{WT}$ using only the sum or mean of each row, calculating the same scores using $T_{RF}$ requires multiplying by the Borda score vector then summing. That is, computing Borda from $T_{RF}$ requires more mathematical operations than computing Borda from $T_{WT}$. Recent work exploring the difficulty of learning basic mathematical operations supports this hypothesis \cite{both2025smallmodels}.  

We find that intuitively simpler rules (Plurality, Borda, and Copeland) are easier to learn, while rules based on iterative procedures that require candidate comparisons or other computations in each iteration, like IRV and Ranked Pairs, are hardest to learn. These more complex rules are also the only rules we study which violate the Monotonicity criterion, suggesting a possible link between Monotonicity and learnability. We also see that rules which are efficiently PAC-learnable, such as Plurality and Borda, are learned more easily than others~\cite{procaccia2009learnability}. 

\paragraph{Single Winner Evaluation}
While our rules are trained to generate a probability distribution over candidates, we can also treat the goal as a single winner prediction task. This allows us to directly compare with the \textit{subset accuracy} metric of \citet{hornischer2025learning}; the winner chosen by a network for a single profile is the candidate receiving the most probability mass. If this is the winner, or one of the tied winners, the output is considered correct. Using 10,000 test profiles we evaluate the accuracy of each trained model on the corresponding voting rule. \Cref{tab:test_acc_all_rules_IC} shows the test accuracy for each rule and each embedding.
While we also learn several rules not studied in prior work, \Cref{tab:comparison_with_others} shows that \textbf{well-chosen embeddings leads to state of the art results using a network with 1-2 orders of magnitude fewer parameters} and a simpler architecture than previous results. 

\begin{table}[tbp]
\caption{
Test accuracy on 10,000 profiles $(m=11, n=44)$ from each preference distribution, for networks trained using Impartial Culture preferences and the $T_{CO}$ embedding. A prediction is correct when the highest-scoring candidate is the winner; these test sets contain no tied winners. Cells show mean $\pm$ standard error of the mean over 5 independently trained models.
}
\centering
\begin{tabular}{@{}lccccc@{}}
\toprule
Target Rule & IAC & Urn & Mall. & SP & PrefL \\ \midrule
\textbf{Plurality} & 1.000 $\pm$ .000 & .989 $\pm$ .002 & .996 $\pm$ .001 & .995 $\pm$ .003 & .973 $\pm$ .002 \\
\textbf{Borda} & .975 $\pm$ .001 & .570 $\pm$ .006 & .477 $\pm$ .006 & .329 $\pm$ .138 & .381 $\pm$ .005 \\
\textbf{Copeland} & 1.000 $\pm$ .000 & 1.000 $\pm$ .000 & 1.000 $\pm$ .000 & 1.000 $\pm$ .000 & .980 $\pm$ .000 \\
\textbf{Schulze} & .917 $\pm$ .004 & .894 $\pm$ .008 & .916 $\pm$ .010 & .957 $\pm$ .015 & .861 $\pm$ .010 \\
\textbf{Simpson-Kramer} & .919 $\pm$ .001 & .898 $\pm$ .005 & .921 $\pm$ .007 & .804 $\pm$ .166 & .866 $\pm$ .009 \\
\textbf{IRV} & .517 $\pm$ .016 & .601 $\pm$ .027 & .652 $\pm$ .033 & .514 $\pm$ .036 & .642 $\pm$ .033 \\
\textbf{Black's Rule} & .915 $\pm$ .002 & .565 $\pm$ .006 & .491 $\pm$ .007 & .560 $\pm$ .127 & .398 $\pm$ .010 \\
\textbf{Ranked Pairs} & .802 $\pm$ .043 & .822 $\pm$ .046 & .854 $\pm$ .049 & .968 $\pm$ .005 & .832 $\pm$ .043 \\ \bottomrule
\end{tabular}
\label{tab:out_of_dist_test_acc}
\end{table}

\paragraph{Beyond Impartial Culture}
\Cref{tab:out_of_dist_test_acc} shows the accuracy of networks trained on $T_{CO}$ tested on 10,000 profiles from several preference distributions, and \Cref{tab:out_of_dist_test_loss} shows the loss of these networks. In most cases, rules trained on $T_{CO}$ are able to generalize effectively to new distributions. On highly structured domains, such as Single-Peaked (SP), some rules (e.g., Black's, Ranked Pairs) have \textit{better} performance than on the Impartial Culture.\footnote{Note that these results are from an experiment using profiles generated by sampling from each preference distribution normally and excluding a profile if it results in tied winners from any voting rule.} Each preference distribution is defined in \Cref{sec:beyond}.

Of the ``simple'' rules, the Copeland rule generalizes almost perfectly across all tested distributions, with Plurality not far behind, while Borda performed surprising poorly outside of the impartial domains (IC and IAC). For all rules other than IRV, either the SP distribution or real-world PrefLib data proved the most challenging, but the shockingly poor generalization of Borda to the SP distribution warrants further investigation. IRV generalizes poorly across the board, which is not surprising considering its performance on IC, but it is surprising that IRV beats Ranked Pairs and Borda on the SP distribution, and its lowest test accuracy was in the IC distribution on which it was trained. Each rule attained similar test accuracy on IAC as it on IC, and performance tended to be similar between Urn and Mallows as well. Of all the theoretical distributions, the Mallows and Urn models showed generalization accuracies closes to the empirical PrefLib data, suggesting that perhaps these model classes correspond more closely to real-world preferences than IC, IAC, or SP.

\textit{The fact that model performance generalizes so well beyond impartial culture is quite surprising at first.} Impartial culture represents a ``uniform random'' of preferences, yet appears highly informative about other preference distributions. This may stem from two factors: As every profile has equal probability of being generated under IC we believe networks effectively learn profiles that are like those which would be sampled from other distributions. Additionally, by focusing learning on settings where many candidates receive similar levels of support the networks are actually learning a \textit{harder} task than if they were to simply learn on a highly structured distribution such as Mallows \cite{mallows1957non}.

\begin{figure*}[tbp]
    \centering
    \includegraphics[width=\textwidth]{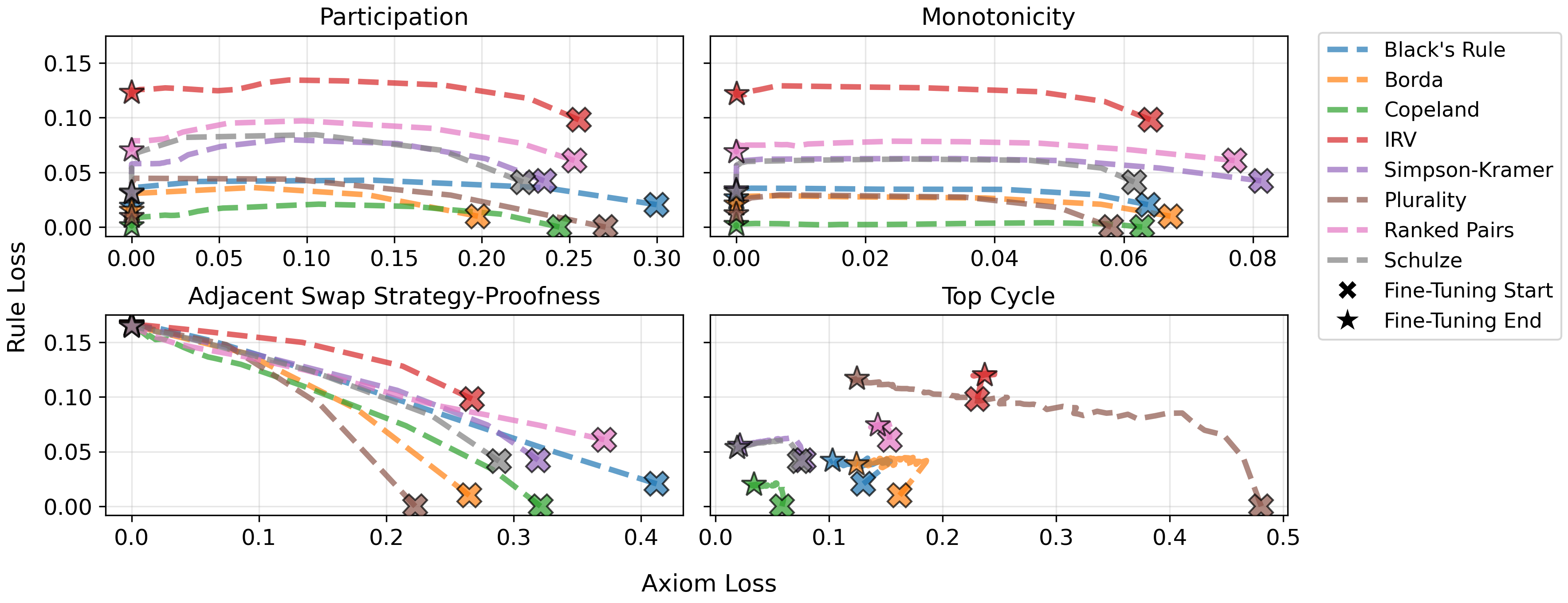}
    \caption{
    Axiom validation loss and rule validation loss during finetuning of models trained on each learned rule using $T_{CO}$ features.
    }
    \label{fig:axiom_loss_rule_loss}
\end{figure*}

\begin{figure*}[tbp]
    \centering
    \includegraphics[width=\linewidth]{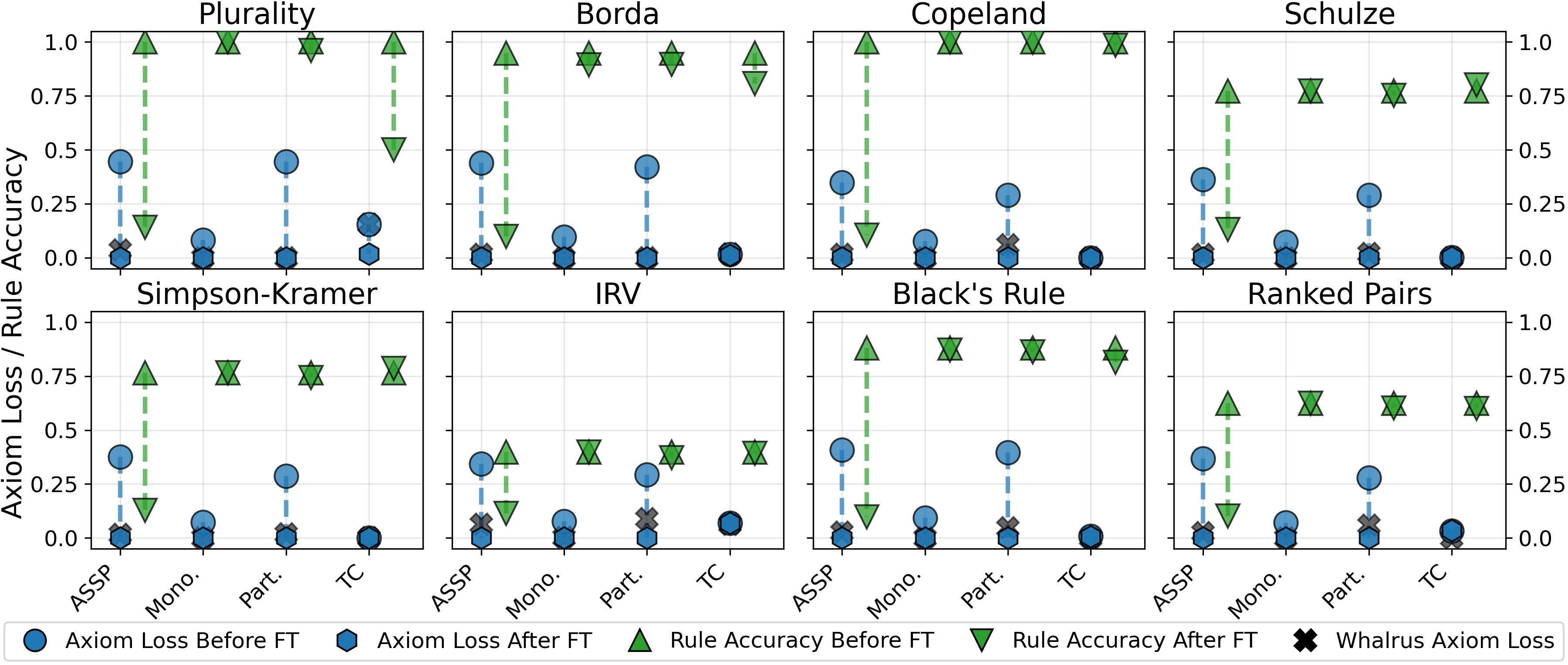}
    \caption{
    Axiom loss and rule accuracy for each rule after training on rule loss (base model) and after finetuning.
    }
    \label{fig:finetune_summary}
\end{figure*}

\section{Fine-Tuning for Axiom Satisfaction}
\label{sec:resist}

We now turn to our primary contribution: using transfer learning to \textit{adapt the existing base models} trained in \Cref{sec:rule_learn} for improved axiomatic performance by applying a second stage of learning using the four axiomatic loss (combined with rule loss) functions we defined in \Cref{sec:axiom_loss_fns}. 

\citet{brandt2024relaxed} show that Condorcet Consistency (which Top Cycle implies) and strong SD-strategyproofness are incompatible for PSCFs. Likewise, \citet{brandt2017optimal} show that Top Cycle and strong SD-participation are also impossible. 
\citet{gibbard1977manipulation} and \citet{sato2013sufficient} show that adjacent strong SD-strategyproofness is equivalent to strong SD-strategyproofness, hence we need only test ASSP to ensure strategyproofness. In fact, Top Cycle mixed with either ASSP or Participation is a \emph{known impossibility} for PSCFs. Hence, in this section we set \emph{impossible voting rules} as our learning target.

\paragraph{Experimental Setup} Fine-tuning uses the same architectures, and profile size as our initial training.
%
We use a batch size of 20 for 100 epochs, training on an NVIDIA A100 GPU for $\approx$ 18 hours per network. Models are trained to minimize the sum of \textit{axiom loss} and \textit{rule loss}, allowing simultaneous propagation of both losses through the network. 
Our inter-profile axioms require exhaustive comparison against $n$ alternative profiles for Participation (each single abstention) and $n(m-1)$ for Monotonicity and Strategyproofness ($m-1$ adjacent swaps for each voter), which becomes computationally prohibitive for even low values of $n$. To mitigate this, we pre-compute a set of \textit{profitable} alternative profiles for each original profile in our new set. A profitable swap or abstention is one which results in a profile that stochastically dominates the original profile. We generate train and test sets of 10,000 profiles each by sampling profiles from the impartial culture distribution and keeping profiles which have a non-zero loss on at least one PSCF and axiom (from ASSP, Participation, or Monotonicity). We restrict our search of adjacent swaps to $k = 5$, for computational tractability and as previous studies have found that human voters nearly always strategize near the top of their preference profile \cite{MWPS15a,scheuerman2021modeling}. This pre-computation greatly reduces the runtime of our loss functions.

\paragraph{Performance of Fine-Tuned PSCFs}

\Cref{fig:axiom_loss_rule_loss} shows how axiom and rule losses change over fine-tuning. \Cref{fig:finetune_summary} gives final axiom loss and rule accuracy on a test set of 10,000 profiles.

Three patterns emerge: \textbf{(1)} Participation and Monotonicity are learned well by all rules. Axiom loss approaches zero with little-to-no increase in rule loss. For rules satisfying Top Cycle, such as Copeland, which is incompatible with Participation, \textbf{these results show a near-impossible outcome}: a version of Copeland which (nearly) satisfies Participation. We find this learned ``Copeland + Participation'' rule primarily differs from Copeland when a voter has a profitable abstention. In such cases, the rule tends to place most probability mass on the alternative that would have been elected in the absence of the potential abstainer. 
On the other hand, \textbf{(2)} ASSP is learned effectively by each rule but at the expense of rule loss: The axiom teaches rules to select a random winner, trivially satisfying strategyproofness. \textbf{(3)} Top Cycle shows additional complexity: 
the Plurality model illustrates the tradeoff between mimicking Plurality and satisfying Top Cycle.
\Cref{app:complete_finetune_results} provides extensive results comparing model performance on each axiom and PSCF before and after fine-tuning.

\section{Learning Axioms Directly}
\label{sec:learn_axioms}

\begin{wraptable}{l}{8.4cm}
\caption{Validation loss for each axiom loss function evaluated on models trained purely on the sum of axiom losses.}
\label{tab:axioms_from_scratch-loss}
\vspace{0.7em}
\centering
\begin{tabular}{@{}lcccc@{}}
\toprule
Axioms              & Part. & Mono. & ASSP  & Top Cycle \\ \midrule
\textbf{ASSP + TC}  & 0     & 0     & 0     & 0.224     \\
\textbf{Part. + TC} & 0     & 0     & 0.448 & 0.12      \\
\textbf{All Axioms} & 0     & 0     & 0     & 0.224     \\ \bottomrule
\end{tabular}
\end{wraptable}

In addition to fine-tuning networks we can also directly train models from scratch for axiom loss. We formed three sets of our axioms, each representing a theoretically impossible combination of axioms. Models are trained for 100 epochs using a loss function equal to the sum of each individual axiom loss function. \Cref{tab:axioms_from_scratch-loss} shows the validation loss after training models on each axiom set. Results are consistent with fine-tuning: Participation and Monotonicity are easy to learn while Strategyproofness and Top Cycle exhibit more complexity. \Cref{app:axioms_from_scratch} shows the training curve of each model and provides additional discussion. We include these results as a proof-of-concept, rather than to exemplify the best or most interesting axiom combinations to learn.

\section{Conclusions and Future Work}


This paper shows that learned voting rules can be imbued with theoretically (almost) impossible axiomatic properties. We first show standard embeddings allow learning existing voting rules with (1) similar or higher accuracy, and (2) a simpler architecture and a fraction of the parameters required by most previous approaches \cite{anil2021learning,hornischer2025learning,li2026learning}. Subsequently, we show that we can fine-tune a trained model to maintain excellence at approximating a learned voting rule while also significantly reducing axiom violations. Different axioms present distinct challenges and come in three varieties. Axioms may be (1) learned well without reducing rule performance (e.g., Participation, Monotonicity),
(2) learned at the cost of rule performance (e.g., ASSP), or (3) difficult to learn in general (e.g., Top Cycle).

These findings suggest that learning social choice axioms directly, in contrast with previous work which has avoided this challenge at the expense of generality \cite{caiata2026what}, will pose novel theoretical challenges to the field. Our results demonstrate that the landscape of axiom loss functions varies dramatically between axioms; popular axioms such as Top Cycle may present a surprising challenge as a learning target. This work can be directly applied to tasks in AI alignment and evaluation which call for novel social choice functions with diverse axiomatic priorities \cite{lanctot2023evaluating,rofin2023vote}.



\clearpage

\section*{Acknowledgements}
Matone, Armstrong, and Mattei were supported in part by NSF Awards IIS-RI-2007955, IIS-III-2107505, IIS-RI-2134857, IIS-RI-2339880 and CNS-SCC-2427237 as well as the Harold L. and Heather E. Jurist Center of Excellence for Artificial Intelligence at Tulane University and the Tulane University Center of Excellence for Community Engaged Artificial Intelligence (CEAI). Ben Abramowitz was supported by the NSF under Grant \#2127309 to the Computing Research Association for the CIFellows Project. Portions of this research were conducted with high performance computational resources provided by the Louisiana Optical Network Infrastructure (LONI) (http://www.loni.org).

\bibliography{refs}


\clearpage
\appendix

\crefalias{section}{appendix}
\crefalias{subsection}{appendix}

\section*{Appendix} 

All code used in this paper is openly available on GitHub: \url{https://github.com/V993/deep-vote}. Authors are happy to discuss additional experiments, future work, or to explain any aspect of the paper.

\section{Voting Rule Definitions}
\label{app:rules}

In this section we give full definitions of each voting rule under study and describe how they are represented as probabilistic social choice functions. We have chosen a wide range of rules to represent those used commonly in practice (e.g., Plurality, IRV) as well as those often studied in social choice settings. Most of the rules function quite differently from one another and have a wide range of informational requirements, providing a strong base from which to explore best practices for learning voting rules. See the Whalrus package documentation for implementation details.

\begin{definition}[Plurality]
    The Plurality score of candidate $c \in C$ from profile $X$ is $L(c) = |\{i \in V : x_i(c) = 1\}|$. Let $W(X) = \argmax\limits_{c \in C} L(c)$ be the subset of candidates with maximum Plurality score. The probabilistic Plurality rule returns the lottery $U(W(X))$.
\end{definition}

\begin{definition}[Borda]
    The Borda score of candidate $c \in C$ from profile $X$ is $B(c) = \sum\limits_{i \in V} (m-x_i(c))$. Let $W(X) = \argmax\limits_{c \in C} B(c)$ be the subset of candidates with maximum Borda score. The probabilistic Borda rule returns the lottery $U(W(X))$ (Referred to as Borda Max by \citet{endriss2017trends}).
\end{definition}

\begin{definition}[Copeland]
    The Copeland score of candidate $c \in C$ from profile $X$ is the number of other candidates it beats in pairwise competition plus $\frac{1}{2}$ times the number of other candidates it ties with in direct competition (if $n$ is even). Let $W(X)$ be the subset of candidates with maximum Copeland score on profile $X$. The probabilistic Copeland rule returns the lottery $U(W(X))$.
\end{definition}

The Simpson-Kramer (Maximin), Schulze, and Ranked Pairs rules are each computed from $T_{WT}$. We will call these weighted-tournament rules. Let $G_X(C,E)$ be the directed tournament graph with edges corresponding to all positive values of the tournament matrix induced by $X$. Let each directed edge $(a,b) \in E$ have weight $d(a,b)=|i \in V : a \succ_i b|$. 

\begin{definition}[Schulze]
    For each path from $a$ to $b$ in $G_X$, we let the strength of the path be the minimum weight edge in that path. For each pair of candidates $a,b \in C$ with a path from $a$ to $b$, we let $p(a,b)$ be the maximum strength of any path from $a$ to $b$, and let $p(a,b)=0$ otherwise.
    Finally, let $W(X) = \{a \in C : p(a,b) \geq p(b,a) \text{ for all } b \in C\}$.
    The probabilistic Schulze rule returns the lottery $U(W(X))$.
\end{definition}

\begin{definition}[Simpson-Kramer]
    Let $W$ be the subset of candidates whose maximum weight incoming edge is minimal in $G_X$. The probabilistic Simpson-Kramer rule returns the lottery $U(W(X))$.
\end{definition}

\begin{definition}[Instant Runoff Voting (IRV)]
    IRV is a deterministic, iterative voting rule that, in each of $m-1$ rounds, eliminates the candidate with the lowest plurality score and removes them from the preference orders of all voters before the next round. When candidates are tied for lowest plurality score we break ties lexicographically. The rule returns the lottery that assigns all probability to the single candidate that was never eliminated; $U(W(X))$ where $|W(X)|=1$.
\end{definition}

\begin{definition}[Black's Rule]
    If the profile $X$ admits a Condorcet winner $c$, then let $W(X) = \{c\}$. Otherwise, if there is no Condorcet winner, let $W(X)$ be the subset of candidates with maximum Borda score $B(c)$. The probabilistic Black's rule returns the lottery $U(W(X))$.
\end{definition}

\begin{definition}[Ranked Pairs]
    Given profile $X$, sort the edges of $G_X$ in descending order by weight, breaking ties lexicographically. Construct an initially empty set $S =\{\}$, and in order by weight, add each edge of $G_X$ to $S$ as long as it would not cause the edges of $S$ to contain a cycle. In the directed acyclic graph defined by $S$, there is necessarily a single undominated candidate $c$ with only edges to other candidates that it beats in pairwise competition. The rule returns a lottery that assigns all probability mass to $c$.
\end{definition}

\section{Preservation Under Embedding}\label{app:preservation}
An embedding may not preserve all of the information needed to compute the output of a function.

\begin{definition}[Function Preservation]
    A function $f : \mathcal{X} \rightarrow \mathcal{Y}$ is preserved by embedding $T: \mathcal{X} \rightarrow \mathcal{X}'$ if $\exists f' : \mathcal{X}' \rightarrow \mathcal{Y}$ such that $f'(T(X)) = f(X)$ for all profiles $X \in \mathcal{X}$.
\end{definition}

Proposition \ref{prop:rule_preserve} says that for an embedding $T$ to preserve a function $f$, there cannot be two profiles with the same embedding under $T$ for which the function returns different outcomes.

\begin{proposition}\label{prop:rule_preserve}
    Embedding $T$ preserves function $f$ if and only if $T(X) = T(\hat{X}) \Rightarrow f(X) = f(\hat{X})$ for all $X, \hat{X} \in \mathcal{X}$.
\end{proposition}

Some embeddings preserve strictly more information than others. 
For instance, $T_{WT}$ preserves all information necessary to compute $T_T$ from a profile. This implies that if $T_T$ preserves a function $f$, then $T_{WT}$ must preserve $f$ as well.

\begin{proposition}\label{prop:intermediate_embedding}
    Suppose that for $T : \mathcal{X} \rightarrow \mathcal{X}'$ there exist $T_1 :  \mathcal{X} \rightarrow \hat{\mathcal{X}}$ and $T_2 :  \hat{\mathcal{X}} \rightarrow \mathcal{X}'$ such that $T(X) = T_2(T_1(X))$ for all $X \in \mathcal{X}$. Then for all $f \in \mathcal{F}$, $T$ preserves $f$ only if $T_1$ preserves $f$.
\end{proposition}

    

Since we can compute $T_T$ from $T_{WT}$, $T_T$ can only preserve a function if $T_{WT}$ does as well. The reverse does not hold, since $T_{WT}$ may preserve functions that are not preserved by $T_T$. If an embedding preserves a function, then the function can be learned to arbitrary accuracy for a given number of candidates using that embedding.

The functions we are concerned with are PSCFs.
Table~\ref{tab:test_loss_all_rules_IC} (green highlights) shows which PSCFs are preserved by these embeddings. See Appendix \ref{app:proofs} 
for proofs of the negative results where PSCFs are not preserved. Each of these proofs consists of a counterexample with two profiles whose outcomes differ under the PSCF yet have the same embedding. 
Finally, many of our rules fall into Fishburn's categorization: the winner of C1 rules (Copeland) can be computed using the information encoded in $T_T$, while the winner of C2 rules (Borda, Schulze, Simpson-Kramer, Black's) can be computed from $T_{WT}$. A separate categorization of voting rules, positional scoring rules (Plurality, Borda), can be computed using only $T_{RF}$ ~\cite{brandt2016handbook}. IRV has recently been shown \textit{not} to belong to either C1 or C2, and is not a positional scoring rule \cite{halpern2024computing}.

\section{Rule Preservation Failure}\label{app:proofs}

Below we provide proofs for each case where we have noted that a PSCF cannot be computed using information found in a specific embedding. Each proof is described by two profiles, $X_1$ and $X_2$. For each embedding-PSCF pair $(T,f)$ below, the profiles $X_1$ and $X_2$ have the same representation in the embedding but the PSCF produces a different lottery; $T(X_1) = T(X_2)$ but $f(X_1) \neq f(X_2)$. This proves that $f$ is not preserved by $T$. For all proofs where the profiles contain 3 voters and 4 candidates, we have confirmed by brute force search that no smaller example exists with 3 voters and 3 candidates.


\begin{table}[ht]
\caption{
PSCF preservation under embedding.
}
\centering
\resizebox{0.35\textwidth}{!}{
\begin{tabular}{|c|c|c|c|}
\hline
& \textbf{$T_{RF}$} & \textbf{$T_{WT}$} & \textbf{$T_T$} \\
\hline
\textbf{Plurality} & \checkmark & $\times$ & $\times$ \\
\hline
\textbf{Borda} & \checkmark & \checkmark & $\times$ \\
\hline
\textbf{Copeland} & $\times$ & \checkmark & \checkmark \\
\hline
\textbf{Schulze} & $\times$ & \checkmark & $\times$ \\
\hline
\textbf{Simpson-Kramer} & $\times$ & \checkmark & $\times$ \\
\hline
\textbf{IRV} & $\times$ & $\times$ & $\times$ \\
\hline
\textbf{Black's Rule} & $\times$ & \checkmark & $\times$ \\
\hline
\textbf{Ranked Pairs} & $\times$ & \checkmark & $\times$ \\
\hline
\end{tabular}
}
\label{tab:rule_transform_preservation}
\end{table}

\subsection{Plurality}
Plurality is a scoring rule, and therefore necessarily computable from a rank frequency embedding. It requires only one column of information from the rank frequency matrix, representing how often each candidate is ranked first by a voter. By contrast, Plurality is not preserved by the weighted tournament embedding, and therefore not by the tournament embedding either.

\begin{theorem}
    The weighted tournament embedding does not preserve Plurality.
\end{theorem}

\begin{proof}
    Let $X_1 = (a \succ b \succ c), (b \succ a \succ c), (c \succ a \succ b)$ and let $X_2 = (a \succ b \succ c), (a \succ b \succ c), (c \succ b \succ a)$. Observe that $f^{Plur}(X_1) = (\frac{1}{3}, \frac{1}{3}, \frac{1}{3})$ because its plurality winner set is $W(X_1) = \{a,b,c\}$, while $f^{Plur}(X_2) = (1, 0, 0)$ because $W(X_2) = \{a\}$ has a unique plurality winner. But in both profiles, there are two voters who prefer $a \succ b$, two voters who prefer $b \succ c$, and two voters who prefer $a \succ c$, so the weighted tournament embeddings are the same $T_{WT}(X_1) = T_{WT}(X_2)$.
    
\end{proof}

\begin{corollary}
    The tournament embedding does not preserve Plurality.
\end{corollary}


\subsection{Borda}

\begin{theorem}
    The tournament embedding does not preserve Borda.
\end{theorem}

\begin{proof}
    $X_1 = (a \succ b \succ c), (a \succ b \succ c), (b \succ c \succ a)$,
    $X_2 = (a \succ b \succ c), (a \succ b \succ c), (a \succ b \succ c)$.
\end{proof}


\subsection{Copeland}

Copeland is the only probabilistic social choice function we consider that is preserved by the tournament embedding, and hence by the weighted tournament as well. However, Copeland is not preserved by the rank frequency embedding.

\begin{theorem}
    The rank frequency embedding does not preserve Copeland.
\end{theorem}

\begin{proof}
    $X_1 = (a \succ b \succ c \succ d), (b \succ c \succ d \succ a), (d \succ a \succ b \succ c)$,
    $X_2 = (a \succ b \succ c \succ d), (b \succ a \succ d \succ c), (d \succ c \succ b \succ a)$.
\end{proof}

Schulze and Simpson-Kramer are weighted-tournament rules that are not preserved by the tournament or rank frequency embedding.

\subsection{Schulze}

\begin{theorem}
    The rank frequency embedding does not preserve Schulze.
\end{theorem}

\begin{proof}
    $X_1 = (a \succ b \succ c \succ d), (b \succ c \succ d \succ a), (d \succ a \succ b \succ c)$,
    $X_2 = (a \succ b \succ c \succ d), (b \succ a \succ d \succ c), (d \succ c \succ b \succ a)$.
\end{proof}

\begin{theorem}
    The tournament embedding does not preserve Schulze.
\end{theorem}

\begin{proof}
    $X_1 = (a \succ b \succ c \succ d), (b \succ c \succ d \succ a), (d \succ a \succ c \succ b)$,
    $X_2 = (a \succ b \succ c \succ d), (b \succ c \succ d \succ a), (d \succ a \succ b \succ c)$.
\end{proof}

\subsection{Simpson-Kramer (Maximin)}

\begin{theorem}
    The rank frequency embedding does not preserve Simpson-Kramer.
\end{theorem}

\begin{proof}
    $X_1 = (a \succ b \succ c \succ d), (b \succ c \succ d \succ a), (d \succ a \succ b \succ c)$,
    $X_2 = (a \succ b \succ c \succ d), (b \succ a \succ d \succ c), (d \succ c \succ b \succ a)$.
\end{proof}

\begin{theorem}
    The tournament embedding does not preserve Simpson-Kramer.
\end{theorem}

\begin{proof}
    $X_1 = (a \succ b \succ c \succ d), (b \succ c \succ a \succ d), (d \succ c \succ a \succ b)$,
    $X_2 = (a \succ b \succ c \succ d), (b \succ c \succ a \succ d), (c \succ a \succ b \succ d)$.
\end{proof}

\subsection{IRV}

\begin{theorem}
    The rank frequency embedding does not preserve IRV.
\end{theorem}

\begin{proof}
    $X_1 = (a \succ b \succ c \succ d), (b \succ c \succ d \succ a), (d \succ a \succ b \succ c)$,
    $X_2 = (a \succ b \succ c \succ d), (b \succ a \succ d \succ c), (d \succ c \succ b \succ a)$. 
\end{proof}

\begin{theorem}
    The tournament embedding does not preserve IRV.
\end{theorem}

\begin{proof}
    $X_1 = (a \succ b \succ c \succ d), (b \succ c \succ d \succ a), (d \succ a \succ c \succ b)$,
    $X_2 = (a \succ b \succ c \succ d), (b \succ c \succ d \succ a), (d \succ a \succ b \succ c)$.
\end{proof}


\begin{theorem}
    The weighted tournament embedding does not preserve IRV.
\end{theorem}
\addtocounter{proof}{1}

This follows as a corollary to \citet{halpern2024computing}. Thanks to Daniel Halpern for pointing this out!

\subsection{Black's Rule}

\begin{theorem}
    The rank frequency embedding does not preserve Black's Rule.
\end{theorem}

\begin{proof}
    $X_1 = (a \succ b \succ c \succ d), (b \succ c \succ d \succ a), (d \succ a \succ b \succ c)$,
    $X_2 = (a \succ b \succ c \succ d), (b \succ a \succ d \succ c), (d \succ c \succ b \succ a)$.
\end{proof}

\begin{theorem}
    The tournament embedding does not preserve Black's Rule.
\end{theorem}

\begin{proof}
    $X_1 = (a \succ b \succ c \succ d), (b \succ c \succ a \succ d), (d \succ c \succ a \succ b)$,
    $X_2 = (a \succ b \succ c \succ d), (b \succ c \succ a \succ d), (c \succ a \succ b \succ d)$.
\end{proof}

The weighted tournament embedding does preserve Black's Rule because it preserves enough information to determine the Condorcet-winner when there is one (e.g., preserves the Copeland Rule), and it preserves the Borda Rule.

\subsection{Ranked Pairs}

\begin{theorem}
    The tournament embedding does not preserve the Ranked Pairs Rule.
\end{theorem}

\begin{proof}
    $X_1 = (a \succ c \succ d \succ b),(c \succ d \succ b \succ a),(d \succ b \succ a \succ c)$, 
    $X_2 = (a \succ c \succ b \succ d),(c \succ d \succ b \succ a),(d \succ b \succ a \succ c)$.
\end{proof}


\begin{theorem}
    The rank frequency embedding does not preserve the Ranked Pairs Rule.
\end{theorem}

\begin{proof}
    $X_1 = (a \succ c \succ b),(b \succ a \succ c),(c \succ b \succ a)$, 
    $X_2 = (a \succ b \succ c),(b \succ c \succ a),(c \succ a \succ b)$.
\end{proof}

\section{Generalization}
\label{app:generalization}

\begin{table}[tbp]
\caption{
L1 loss on the validation split of the training data (Impartial Culture, $(m=11, n=44)$) at the epoch with the lowest validation loss, for models targeting each rule using each embedding. Cells show mean $\pm$ standard error of the mean over 5 independently trained models. Shaded cells indicate the embedding contains sufficient information to learn the rule perfectly.
}
\centering
\begin{tabular}{@{}lcccc@{}}
\toprule
\textbf{} & $T_{RF}$ & $T_{T}$ & $T_{WT}$ & $T_{CO}$ \\ \midrule
\textbf{Plurality} & \cellcolor{green!25}0.003 $\pm$ 0.003 & 0.127 $\pm$ 0.000 & 0.126 $\pm$ 0.000 & \cellcolor{green!25}0.000 $\pm$ 0.000 \\
\textbf{Borda} & \cellcolor{green!25}0.036 $\pm$ 0.005 & 0.047 $\pm$ 0.000 & \cellcolor{green!25}0.022 $\pm$ 0.003 & \cellcolor{green!25}0.009 $\pm$ 0.000 \\
\textbf{Copeland} & 0.072 $\pm$ 0.003 & \cellcolor{green!25}0.000 $\pm$ 0.000 & \cellcolor{green!25}0.055 $\pm$ 0.002 & \cellcolor{green!25}0.000 $\pm$ 0.000 \\
\textbf{Schulze} & 0.066 $\pm$ 0.001 & 0.040 $\pm$ 0.000 & \cellcolor{green!25}0.061 $\pm$ 0.000 & \cellcolor{green!25}0.040 $\pm$ 0.000 \\
\textbf{Simpson-Kramer} & 0.066 $\pm$ 0.001 & 0.044 $\pm$ 0.002 & \cellcolor{green!25}0.063 $\pm$ 0.001 & \cellcolor{green!25}0.041 $\pm$ 0.000 \\
\textbf{IRV} & 0.109 $\pm$ 0.002 & 0.101 $\pm$ 0.002 & 0.105 $\pm$ 0.002 & 0.096 $\pm$ 0.002 \\
\textbf{Black's Rule} & 0.043 $\pm$ 0.003 & 0.043 $\pm$ 0.000 & \cellcolor{green!25}0.020 $\pm$ 0.003 & \cellcolor{green!25}0.022 $\pm$ 0.001 \\
\textbf{Ranked Pairs} & 0.089 $\pm$ 0.003 & 0.047 $\pm$ 0.004 & \cellcolor{green!25}0.078 $\pm$ 0.002 & \cellcolor{green!25}0.052 $\pm$ 0.005 \\ \bottomrule
\end{tabular}
\label{tab:test_loss_all_rules_IC}
\end{table}

Our embeddings give us the ability to work with profiles with different numbers of voters. Although we trained our models on profiles with 44 voters, we can test with larger and smaller numbers of voters to assess generalization. The following experiments are performed on preliminary models which are not identical to the main results in this paper. The performance of these models is close to those in our final report. We test first on profiles with 199 votes (\Cref{tab:rule_loss_199v11c}), and then again with only 13 voters (\Cref{tab:rule_loss_13v11c}). 

We see that there is a very mild increase in the loss across most rules and embeddings, though losses remain very similar to their value on profiles with 44 voters (see \Cref{tab:test_loss_all_rules_IC}). In some cases there is no increase in loss: With the concatenated embedding, $T_{CO}$, both Copeland (with 13 voters) and Black's rule (199 voters) have the same loss. Loss even decreases when evaluating Black's rule on profiles with 13 voters and $T_{T}$. 
As well, on all but 4 rule-embedding pairs, loss is lower using profiles with 13 voters than those with 199 voters. 
These results highlight the importance of choosing embeddings that fit the rule, corresponding to the learning objective. As a result of our embeddings our learned rules are able to generalize extremely effectively to profiles of varying sizes without the need for additional training.


\begin{table}[tbp]
\caption{
L1 loss on 512 Impartial Culture profiles with 11 candidates and 199 voters, for networks trained on each embedding with 11 candidates and 44 voters. Cells show mean $\pm$ standard error of the mean over 5 independently trained models.
}
\centering
\begin{tabular}{@{}lcccc@{}}
\toprule
Target Rule & $T_{RF}$ & $T_{T}$ & $T_{WT}$ & $T_{CO}$ \\ \midrule
\textbf{Plurality} & 0.041 $\pm$ 0.005 & 0.134 $\pm$ 0.000 & 0.135 $\pm$ 0.001 & 0.043 $\pm$ 0.005 \\
\textbf{Borda} & 0.039 $\pm$ 0.004 & 0.060 $\pm$ 0.001 & 0.025 $\pm$ 0.003 & 0.025 $\pm$ 0.001 \\
\textbf{Copeland} & 0.084 $\pm$ 0.004 & 0.000 $\pm$ 0.000 & 0.072 $\pm$ 0.001 & 0.000 $\pm$ 0.000 \\
\textbf{Schulze} & 0.075 $\pm$ 0.003 & 0.056 $\pm$ 0.001 & 0.069 $\pm$ 0.002 & 0.055 $\pm$ 0.001 \\
\textbf{Simpson-Kramer} & 0.071 $\pm$ 0.002 & 0.058 $\pm$ 0.001 & 0.070 $\pm$ 0.001 & 0.056 $\pm$ 0.001 \\
\textbf{IRV} & 0.118 $\pm$ 0.003 & 0.104 $\pm$ 0.003 & 0.111 $\pm$ 0.002 & 0.102 $\pm$ 0.002 \\
\textbf{Black's Rule} & 0.049 $\pm$ 0.003 & 0.049 $\pm$ 0.001 & 0.028 $\pm$ 0.003 & 0.037 $\pm$ 0.001 \\
\textbf{Ranked Pairs} & 0.097 $\pm$ 0.003 & 0.055 $\pm$ 0.004 & 0.082 $\pm$ 0.002 & 0.063 $\pm$ 0.006 \\ \bottomrule
\end{tabular}
\label{tab:rule_loss_199v11c}
\end{table}


\begin{table}[tbp]
\caption{
L1 loss on 512 Impartial Culture profiles with 11 candidates and 13 voters, for networks trained on each embedding with 11 candidates and 44 voters. Cells show mean $\pm$ standard error of the mean over 5 independently trained models.
}
\centering
\begin{tabular}{@{}lcccc@{}}
\toprule
Target Rule & $T_{RF}$ & $T_{T}$ & $T_{WT}$ & $T_{CO}$ \\ \midrule
\textbf{Plurality} & 0.004 $\pm$ 0.003 & 0.119 $\pm$ 0.001 & 0.117 $\pm$ 0.000 & 0.005 $\pm$ 0.001 \\
\textbf{Borda} & 0.037 $\pm$ 0.006 & 0.053 $\pm$ 0.001 & 0.025 $\pm$ 0.003 & 0.023 $\pm$ 0.000 \\
\textbf{Copeland} & 0.076 $\pm$ 0.002 & 0.000 $\pm$ 0.000 & 0.060 $\pm$ 0.001 & 0.000 $\pm$ 0.000 \\
\textbf{Schulze} & 0.068 $\pm$ 0.001 & 0.048 $\pm$ 0.001 & 0.064 $\pm$ 0.001 & 0.048 $\pm$ 0.000 \\
\textbf{Simpson-Kramer} & 0.067 $\pm$ 0.001 & 0.051 $\pm$ 0.002 & 0.064 $\pm$ 0.001 & 0.048 $\pm$ 0.001 \\
\textbf{IRV} & 0.109 $\pm$ 0.002 & 0.103 $\pm$ 0.003 & 0.110 $\pm$ 0.002 & 0.094 $\pm$ 0.002 \\
\textbf{Black's Rule} & 0.049 $\pm$ 0.003 & 0.040 $\pm$ 0.001 & 0.031 $\pm$ 0.003 & 0.028 $\pm$ 0.001 \\
\textbf{Ranked Pairs} & 0.093 $\pm$ 0.005 & 0.046 $\pm$ 0.004 & 0.077 $\pm$ 0.002 & 0.051 $\pm$ 0.006 \\ \bottomrule
\end{tabular}
\label{tab:rule_loss_13v11c}
\end{table}

\subsection{Beyond Impartial Culture}
\label{sec:beyond}

Recent \cite{boehmer2024guide,boehmer2022expected,szufa2025drawing} and older work \cite{mattei2013preflib,mattei2017apreflib} in the COMSOC community has illustrated the need to test on a wide variety of synthetic and real world preference distributions to ensure generalization. We test the empirical merit of this fact by evaluating our networks trained on IC preferences on test sets sampled from a wide range of distributions, including real-world preference data. 
In \Cref{sec:rule_learn} we show how well rules trained on Impartial Culture preferences generalize to other preference distributions. The main body shows the accuracy of each rule and \Cref{tab:out_of_dist_test_loss} shows the test loss for each model across distributions.
Specifically, we consider the following distributions:

\begin{itemize}
    \item[] \textbf{Impartial Culture} (IC)  Each unique preference order is equally likely, regardless of which orders any other voters have selected ~\cite{guilbaud1952theories}.
    
    \item[] \textbf{Impartial Anonymous Culture} (IAC) 
    Preferences profiles are generated collectively, rather than as individual preference orders.
    Each multi-set of preference orders (i.e. a preference profile) is equally likely to be generated, making voter identities irrelevant
    ~\cite{gehrlein1976condorcet}.

    \item[] \textbf{Mallows} Preference orders are noisy estimates of some reference ranking $r$, with the amount of noise related to a parameter $\phi$ ~\cite{mallows1957non}. A value of $\phi = 0$ results in all voters have identical preferences while $\phi = 1$ results in the Impartial Culture distribution. For each profile we sample $\phi \in [0, 1]$ uniformly at random using the Mallow's distribution described by ~\citet{boehmer2021putting}.
    
    \item[] \textbf{Urn} All $m!$ preference orders exist in an ``urn.'' Each voter decides their ranking by sampling a ranking from the urn. Once a ranking is selected, $\alpha m!$ copies of it are added to the urn ~\cite{eggenberger1923statistik}. For each profile we sample $\alpha$ from a Gamma distribution with shape parameter $k = 0.8$ and scale parameter $\theta = 1$ as described by \citet{boehmer2021putting}.
    
    \item[] \textbf{Single-Peaked} 
    There is some global ordering of alternatives. Each voter has some favourite alternative and prefers all alternatives closer to their favourite over those further away. We sample single-peaked profiles from Walsh's distribution \cite{walsh2015generating}.

    \item[] \textbf{PrefLib} An online repository containing profiles corresponding to real human preferences expressed across many domains ~\cite{mattei2013preflib}. We use all complete profiles with strict orders and $m \geq 11$ candidates. For profiles with greater than $11$ candidates we form a profile by selecting a subset of candidates uniformly at random. This results in a test set of 6945 profiles.
\end{itemize}

\begin{table}[tbp]
\caption{
Test loss on 10,000 profiles $(m=11, n=44)$ from each preference distribution, for networks trained using Impartial Culture preferences and the $T_{CO}$ embedding. Cells show mean $\pm$ standard error of the mean over 5 independently trained models.}
\centering
\resizebox{\textwidth}{!}{%
\begin{tabular}{@{}lcccccc@{}}
\toprule
Target Rule & IC & IAC & Urn & Mall. & SP & PrefLib \\ \midrule
\textbf{Plurality} & .000 $\pm$ .000 & .000 $\pm$ .000 & .005 $\pm$ .000 & .002 $\pm$ .000 & .006 $\pm$ .001 & .003 $\pm$ .000 \\
\textbf{Borda} & .005 $\pm$ .000 & .004 $\pm$ .000 & .078 $\pm$ .001 & .095 $\pm$ .001 & .123 $\pm$ .025 & .112 $\pm$ .001 \\
\textbf{Copeland} & .000 $\pm$ .000 & .000 $\pm$ .000 & .000 $\pm$ .000 & .000 $\pm$ .000 & .000 $\pm$ .000 & .000 $\pm$ .000 \\
\textbf{Schulze} & .017 $\pm$ .000 & .017 $\pm$ .000 & .024 $\pm$ .001 & .023 $\pm$ .001 & .026 $\pm$ .011 & .032 $\pm$ .001 \\
\textbf{SK} & .017 $\pm$ .000 & .016 $\pm$ .000 & .024 $\pm$ .001 & .022 $\pm$ .001 & .040 $\pm$ .020 & .032 $\pm$ .001 \\
\textbf{IRV} & .088 $\pm$ .003 & .088 $\pm$ .003 & .073 $\pm$ .005 & .063 $\pm$ .006 & .088 $\pm$ .007 & .065 $\pm$ .006 \\
\textbf{Black's} & .014 $\pm$ .000 & .015 $\pm$ .000 & .079 $\pm$ .001 & .093 $\pm$ .001 & .080 $\pm$ .023 & .110 $\pm$ .002 \\
\textbf{RP} & .036 $\pm$ .008 & .036 $\pm$ .008 & .032 $\pm$ .008 & .026 $\pm$ .009 & .006 $\pm$ .001 & .030 $\pm$ .008 \\
\bottomrule
\end{tabular}%
}
\label{tab:out_of_dist_test_loss}
\end{table}

\section{Other Tested Neural Network Setups}\label{app:other_test}

We note that in all cases, our results are very similar between runs using the same parameters. Across different parameter settings our results are broadly similar -- rules that are learned well are learned well and rules learned less well have worse outcomes -- in all settings.

Before we settled on our final experiment configuration, we conducted several tests to explore the efficiency of learning and target larger profile dimensionalities without making significant changes to our models or embeddings. Experimenting with our model’s representational threshold, we found that with higher dimensionalities from our initial baseline of 7 candidates and 29 voters we were able to achieve faster and more consistent convergence with L1Loss and TanH activation.

With 15 candidates and 44 voters, we observe training loss decrease much more consistently than our ReLU baseline with lower variance and better generalization. Similarly, we find that utilization of a distribution-based loss like KL-Divergence or Jensen-Shannon divergence is inefficient when targeting a sparse output vector, and that even with 7 candidates these approaches see a severe reduction in the generalizability of our models.

While we did want to focus on smaller model architectures to more directly rely on our embeddings, we did experiment with more complex models. We tested two alternate networks which we observed to have below-satisfactory performance when compared to our base feed-forward network. The first was a simple network with a depth of 6 and a constant layer width of $m^2 \times 3$, 
and the second had the same depth, but a funnel architecture which decomposed the width of each hidden layer from the input to the output layer to minimize information dropoff. Both networks underperformed when compared to our baseline.

\section{Learning Axioms Directly}
\label{app:axioms_from_scratch}

\begin{figure}[htbp]
    \centering
    \includegraphics[width=0.6\linewidth]{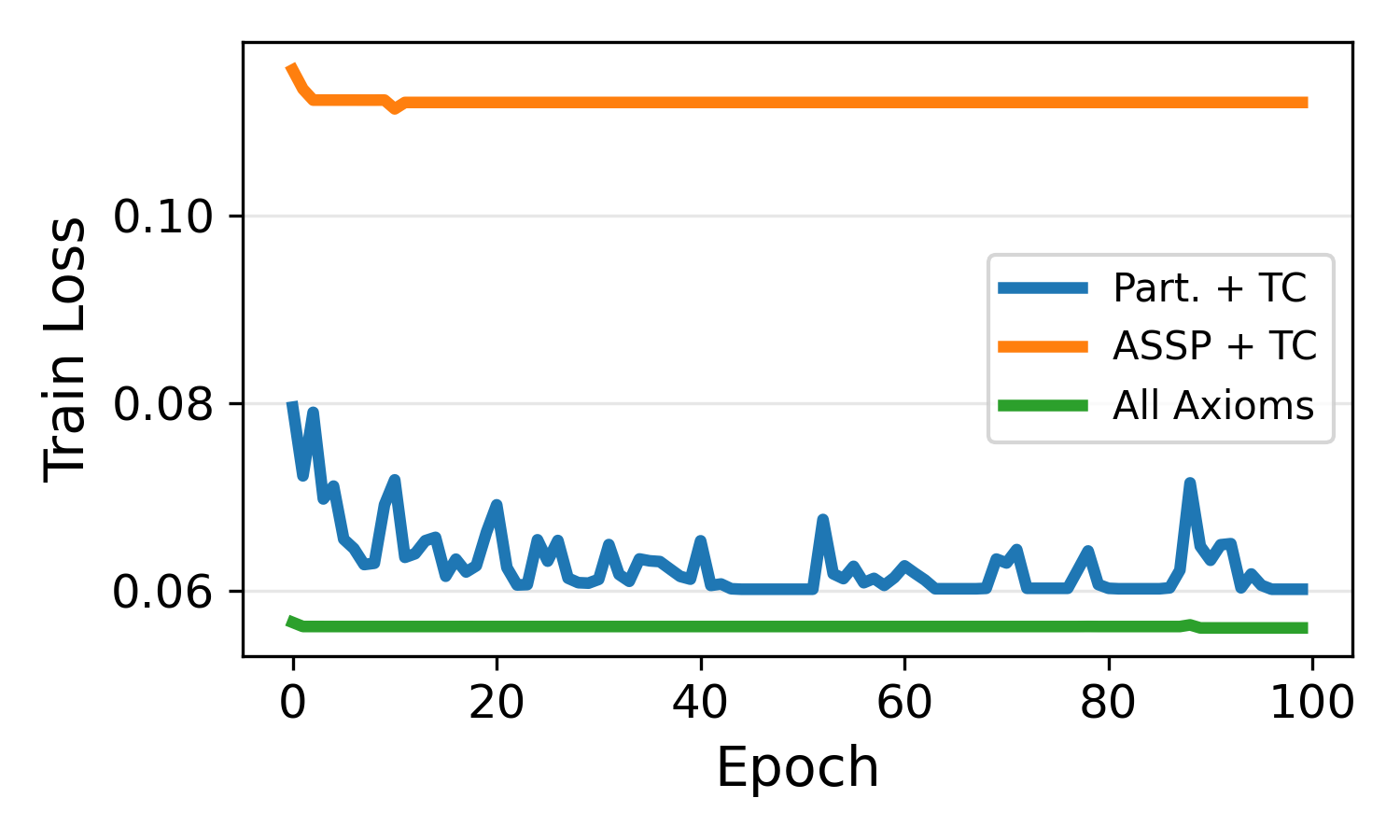}
    \caption{Training loss of each set of axioms on networks trained exclusively for axioms loss.}
    \label{fig:axioms_from_scratch-loss_curves}
\end{figure}


Here we provide results and additional discussion around our experiments training networks purely focused on learning axioms. We fix three subsets of axioms: (1) Strategyproofness and Top Cycle \cite{brandt2024relaxed}, (2) Participation and Top Cycle \cite{moulin1988condorcet}, and (3) all four axioms. Each of these axiom represents a known impossibility. Using the same architecture as our other experiments we train a network targeting a loss function equal to the unweighted sum of each axiom's individual loss function.

\Cref{fig:axioms_from_scratch-loss_curves} shows the training loss of each network over 100 epochs of training. Final axiom loss values for each axiom are shown in \Cref{tab:axioms_from_scratch-loss}. We observe that loss only decreases very little and expect this is caused by a highly discrete loss landscape -- a function must change significantly in order to reduce loss. Similar experiments (not shown) with a higher learning rate did not fundamentally change this result. While the axioms are, in each case, mutually incompatible and reaching a loss of zero is impossible we suspect different learning techniques may be more well suited for this task and provide superior results.

A particularly striking aspect of \Cref{tab:axioms_from_scratch-loss} is that Participation and Monotonicity always have approximately 0 loss. Participation's lack of loss can be explained by the fact that an untrained randomized network will never violate the Participation axiom; while we do not prove it we expect that learning to minimize ASSP loss also adds pressure to reduce Participation loss (otherwise we would expect ASSP + TC to increase Participation loss).
Monotonicity appears a relatively ``easy'' axiom to avoid violating and is closely tied to ASSP.
We also note that the high loss found for Top Cycle is conceptually similar to the result of \citet{caiata2026what} (under a very different model) who also found learning rules which avoid violating the Condorcet property to be challenging.

\section{Complete Axiom Finetuning Results}
\label{app:complete_finetune_results}

\Cref{tab:MonotonicityLoss-axiom_loss,tab:MonotonicityLoss-axiom_violation_rate,tab:MonotonicityLoss-rule_loss,tab:ParticipationLoss-axiom_loss,tab:ParticipationLoss-axiom_violation_rate,tab:ParticipationLoss-rule_loss,tab:TopCycleLoss-axiom_loss,tab:TopCycleLoss-axiom_violation_rate,tab:TopCycleLoss-rule_loss,tab:AsspLoss-axiom_loss,tab:AsspLoss-axiom_violation_rate,tab:AsspLoss-rule_loss} report axiom loss, rule loss, and axiom violation rate for each rule and each type of rule (actual rule, base model trained for rule loss, finetuned model). All trained models used $T_{CO}$ features. An axiom is said to have been violated on an example if the axiom loss on the example is greater than $1e^{-5}$. We report results for a single training run only. Our other results show that the variability between individual training runs is quite small.

\begin{table*}[htbp!]
\caption{\underline{Axiom Loss} of all trained models and actual rules on \underline{Monotonicity} axiom.}
\begin{tabular}{lcccccccc}
\toprule
 & Plurality & Borda & Copeland & Schulze & SK              & IRV & Black's      & RP           \\
\midrule
Base Model & 0.083 & 0.096 & 0.076 & 0.072 & 0.073 & 0.077 & 0.094 & 0.071 \\
Finetuned       & 0.000 & 0.000 & 0.000 & 0.000 & 0.000 & 0.000 & 0.000 & 0.000 \\
Rule & 0.000 & 0.000 & 0.000 & 0.000 & 0.000 & 0.003 & 0.000 & 0.001 \\
\bottomrule
\end{tabular}
\label{tab:MonotonicityLoss-axiom_loss}
\end{table*}


\begin{table*}[htbp!]
\caption{\underline{Axiom Violation Rate} of all trained models and actual rules on \underline{Monotonicity} axiom.}
\begin{tabular}{lcccccccc}
\toprule
 & Plurality & Borda & Copeland & Schulze & SK              & IRV & Black's      & RP           \\
\midrule
Base Model & 0.842 & 0.822 & 0.742 & 0.866 & 0.839 & 0.624 & 0.808 & 0.597 \\
Finetuned       & 0.029 & 0.003 & 0.010 & 0.016 & 0.009 & 0.000 & 0.001 & 0.000 \\
Rule & 0.000 & 0.000 & 0.000 & 0.000 & 0.000 & 0.166 & 0.000 & 0.026 \\
\bottomrule
\end{tabular}
\label{tab:MonotonicityLoss-axiom_violation_rate}
\end{table*}


\begin{table*}[htbp!]
\caption{\underline{Rule Loss} of all trained models and actual rules on \underline{Monotonicity} axiom.}
\begin{tabular}{lcccccccc}
\toprule
 & Plurality & Borda & Copeland & Schulze & SK              & IRV & Black's      & RP           \\
\midrule
Base Model & 0.000 & 0.013 & 0.000 & 0.054 & 0.056 & 0.119 & 0.026 & 0.069 \\
Finetuned       & 0.004 & 0.027 & 0.004 & 0.057 & 0.059 & 0.121 & 0.031 & 0.072 \\
Rule & 0.000 & 0.000 & 0.000 & 0.000 & 0.000 & 0.000 & 0.000 & 0.000 \\
\bottomrule
\end{tabular}
\label{tab:MonotonicityLoss-rule_loss}
\end{table*}


\begin{table*}[htbp!]
\caption{\underline{Axiom Loss} of all trained models and actual rules on \underline{Participation} axiom.}
\begin{tabular}{lcccccccc}
\toprule
 & Plurality & Borda & Copeland & Schulze & SK              & IRV & Black's      & RP           \\
\midrule
Base Model & 0.446 & 0.422 & 0.291 & 0.291 & 0.288 & 0.293 & 0.396 & 0.280 \\
Finetuned       & 0.000 & 0.000 & 0.000 & 0.000 & 0.000 & 0.000 & 0.000 & 0.000 \\
Rule & 0.000 & 0.000 & 0.060 & 0.019 & 0.019 & 0.090 & 0.047 & 0.059 \\
\bottomrule
\end{tabular}
\label{tab:ParticipationLoss-axiom_loss}
\end{table*}


\begin{table*}[htbp!]
\caption{\underline{Axiom Violation Rate} of all trained models and actual rules on \underline{Participation} axiom.}
\begin{tabular}{lcccccccc}
\toprule
 & Plurality & Borda & Copeland & Schulze & SK              & IRV & Black's      & RP           \\
\midrule
Base Model & 0.944 & 0.890 & 0.783 & 0.911 & 0.879 & 0.656 & 0.880 & 0.641 \\
Finetuned       & 0.052 & 0.001 & 0.003 & 0.006 & 0.014 & 0.000 & 0.000 & 0.000 \\
Rule & 0.000 & 0.000 & 0.405 & 0.257 & 0.271 & 0.330 & 0.100 & 0.215 \\
\bottomrule
\end{tabular}
\label{tab:ParticipationLoss-axiom_violation_rate}
\end{table*}


\begin{table*}[htbp!]
\caption{\underline{Rule Loss} of all trained models and actual rules on \underline{Participation} axiom.}
\begin{tabular}{lcccccccc}
\toprule
 & Plurality & Borda & Copeland & Schulze & SK              & IRV & Black's      & RP           \\
\midrule
Base Model & 0.000 & 0.013 & 0.000 & 0.054 & 0.056 & 0.119 & 0.026 & 0.069 \\
Finetuned       & 0.017 & 0.023 & 0.003 & 0.058 & 0.059 & 0.123 & 0.029 & 0.073 \\
Rule & 0.000 & 0.000 & 0.000 & 0.000 & 0.000 & 0.000 & 0.000 & 0.000 \\
\bottomrule
\end{tabular}
\label{tab:ParticipationLoss-rule_loss}
\end{table*}


\begin{table*}[htbp!]
\caption{\underline{Axiom Loss} of all trained models and actual rules on \underline{Top Cycle} axiom.}
\begin{tabular}{lcccccccc}
\toprule
 & Plurality & Borda & Copeland & Schulze & SK              & IRV & Black's      & RP           \\
\midrule
Base Model & 0.155 & 0.015 & 0.000 & 0.001 & 0.001 & 0.069 & 0.009 & 0.032 \\
Finetuned       & 0.017 & 0.012 & 0.000 & 0.001 & 0.001 & 0.070 & 0.007 & 0.035 \\
Rule & 0.155 & 0.017 & 0.000 & 0.000 & 0.000 & 0.063 & 0.001 & 0.000 \\
\bottomrule
\end{tabular}
\label{tab:TopCycleLoss-axiom_loss}
\end{table*}


\begin{table*}[htbp!]
\caption{\underline{Axiom Violation Rate} of all trained models and actual rules on \underline{Top Cycle} axiom.}
\begin{tabular}{lcccccccc}
\toprule
 & Plurality & Borda & Copeland & Schulze & SK              & IRV & Black's      & RP           \\
\midrule
Base Model & 0.210 & 0.017 & 0.014 & 0.009 & 0.006 & 0.069 & 0.011 & 0.032 \\
Finetuned       & 0.060 & 0.013 & 0.050 & 0.004 & 0.012 & 0.070 & 0.008 & 0.035 \\
Rule & 0.192 & 0.018 & 0.000 & 0.000 & 0.000 & 0.063 & 0.001 & 0.000 \\
\bottomrule
\end{tabular}
\label{tab:TopCycleLoss-axiom_violation_rate}
\end{table*}


\begin{table*}[htbp!]
\caption{\underline{Rule Loss} of all trained models and actual rules on \underline{Top Cycle} axiom.}
\begin{tabular}{lcccccccc}
\toprule
 & Plurality & Borda & Copeland & Schulze & SK              & IRV & Black's      & RP           \\
\midrule
Base Model & 0.000 & 0.013 & 0.000 & 0.054 & 0.056 & 0.119 & 0.026 & 0.069 \\
Finetuned       & 0.115 & 0.039 & 0.018 & 0.060 & 0.063 & 0.120 & 0.038 & 0.073 \\
Rule & 0.000 & 0.000 & 0.000 & 0.000 & 0.000 & 0.000 & 0.000 & 0.000 \\
\bottomrule
\end{tabular}
\label{tab:TopCycleLoss-rule_loss}
\end{table*}


\begin{table*}[htbp!]
\caption{\underline{Axiom Loss} of all trained models and actual rules on \underline{ASSP} axiom.}
\begin{tabular}{lcccccccc}
\toprule
 & Plurality & Borda & Copeland & Schulze & SK              & IRV & Black's      & RP           \\
\midrule
Base Model & 0.446 & 0.440 & 0.348 & 0.363 & 0.377 & 0.343 & 0.409 & 0.369 \\
Finetuned       & 0.000 & 0.000 & 0.000 & 0.000 & 0.000 & 0.000 & 0.000 & 0.000 \\
Rule & 0.036 & 0.015 & 0.016 & 0.015 & 0.016 & 0.064 & 0.027 & 0.025 \\
\bottomrule
\end{tabular}
\label{tab:AsspLoss-axiom_loss}
\end{table*}


\begin{table*}[htbp!]
\caption{\underline{Axiom Violation Rate} of all trained models and actual rules on \underline{ASSP} axiom.}
\begin{tabular}{lcccccccc}
\toprule
 & Plurality & Borda & Copeland & Schulze & SK              & IRV & Black's      & RP           \\
\midrule
Base Model & 0.964 & 0.828 & 0.825 & 0.914 & 0.912 & 0.678 & 0.805 & 0.744 \\
Finetuned       & 0.000 & 0.000 & 0.000 & 0.000 & 0.000 & 0.000 & 0.000 & 0.000 \\
Rule & 0.678 & 0.131 & 0.404 & 0.512 & 0.516 & 0.565 & 0.169 & 0.329 \\
\bottomrule
\end{tabular}
\label{tab:AsspLoss-axiom_violation_rate}
\end{table*}


\begin{table*}[htbp!]
\caption{\underline{Rule Loss} of all trained models and actual rules on \underline{ASSP} axiom.}
\begin{tabular}{lcccccccc}
\toprule
 & Plurality & Borda & Copeland & Schulze & SK              & IRV & Black's      & RP           \\
\midrule
Base Model & 0.000 & 0.013 & 0.000 & 0.054 & 0.056 & 0.119 & 0.026 & 0.069 \\
Finetuned       & 0.165 & 0.165 & 0.166 & 0.165 & 0.165 & 0.165 & 0.165 & 0.164 \\
Rule & 0.000 & 0.000 & 0.000 & 0.000 & 0.000 & 0.000 & 0.000 & 0.000 \\
\bottomrule
\end{tabular}
\label{tab:AsspLoss-rule_loss}
\end{table*}

\end{document}